\documentclass[journal]{IEEEtran}

\usepackage{color}
\usepackage{bbm}
\usepackage{amssymb}
\usepackage{amsmath}
\usepackage{amsfonts}
\usepackage{subeqnarray}
\usepackage{cases}
\usepackage{multicol}
\usepackage{lipsum}
\usepackage{array}
\usepackage{bm}
\usepackage{graphicx}

\usepackage{algorithm}
\usepackage{algorithmic}

\newcommand{\RNum}[1]{\uppercase\expandafter{\romannumeral #1\relax}}

\title{Adaptive Offline and Online Similarity-Based Caching}

\begin{document}

\author{Jizhe~Zhou,~Osvaldo~Simeone,~Xing~Zhang,~and~Wenbo~Wang
\thanks{J. Zhou, X. Zhang and W. Wang are with School of Information and Communications Engineering, Beijing University of Posts and Telecommunications, Beijing 100876, China (\emph{Corresponding author: Xing Zhang}, email: zhangx@ieee.org).}
\thanks{O. Simeone is with the Department of Engineering, King’s College
London, London WC2R 2LS, U.K (email: osvaldo.simeone@kcl.ac.uk).}
\thanks{The work of J. Zhou, X. Zhang and W. Wang was supported by the National Science Foundation of China under Grant 61771065, 62071063 and 61631005 in part by the Zhejiang Laboratory Open Project Fund 2020LCOAB01. The work of J. Zhou was supported by China Scholarship Council. The work of O. Simeone was supported by the European Research Council (ERC) under the European Union’s Horizon 2020 Research and Innovation Programme (Grant Agreement No. 725731).}
}

\maketitle

\begin{abstract}
  With similarity-based content delivery, the request for a content can be satisfied by delivering a related content under a dissimilarity cost. This letter addresses the joint optimization of caching and similarity-based delivery decisions across a network so as to minimize the weighted sum of average delay and dissimilarity cost. A convergent alternate gradient descent ascent algorithm is first introduced for an offline scenario with prior knowledge of the request rates, and then extended to an online setting. Numerical results validate the advantages of the approach with respect to standard per-cache solutions.
\end{abstract}


\section{Introduction}

\IEEEPARstart{C}{aching} systems provide the underlying architecture for content-centric networks \cite{1}, content distribution networks \cite{2}, and edge networks \cite{3}. In conventional systems, a request for a content is satisfied by forwarding it to a node that permanently stores the requested content. Caching networks can reduce the delivery delay by serving the request from one of the intermediate nodes in the forwarding path that stores the requested content (see Fig. 1).  

In many applications, a request can also be satisfied by delivering a content similar to the requested one \cite{4}. Examples include video and image retrieval as well as advertising \cite{5,6}. For example, when a user searches for a video, a related video cached locally may be delivered instead, as long as the resulting lower downloading latency offsets the ``dissimilarity cost" associated with receiving a different content. 

\begin{figure}
    \centering
    \includegraphics[width=0.4\textwidth]{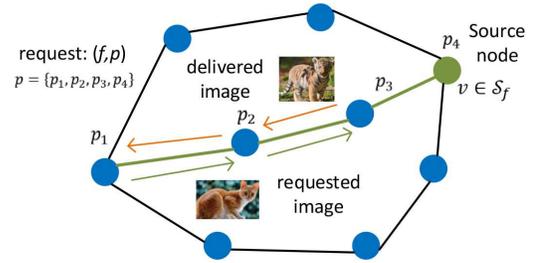}
    \caption{Similarity-based delivery: The green line depicts the forwarding path $p$ for the requested content $f$. While the request is for a ``cat'' image (as depicted over the green arrowed line), a similar image of a ``tiger cub'' cached at node $p_3\in p$ is delivered instead (as depicted over the orange arrowed line).}
    \label{fig1}
\end{figure}

Motivated by these considerations, we study similarity-based content caching and delivery in a cache-enabled network. As illustrated in Fig. 1, a request for a content is routed over a path to a designated node that permanently stores the requested content. Ioannidis and Yeh \cite{10} studied the conventional case in which the request is satisfied by delivering the requested content from one of the caches along the path, if a ``cache hit" occurs, or from the end node otherwise. In contrast, in this paper, we allow for similarity-based delivery. Accordingly, a similar content can be delivered if it is found in one of the caches along the path -- an event known as ``soft cache hit" \cite{7,8}. To the best of our knowledge, prior work on similarity caching focuses on \emph{per-cache} strategies that deliver the most similar content from a fixed local cache for each request \cite{4,5}. As in \cite{10}, we allow instead for coordination across the caches in the network, and consider the joint optimization of caching and delivery decisions, where a hit can occur at any of the caches along a path for a request. 

Specifically, we first study the offline optimization problem over cache allocation and delivery decisions, such that the weighted sum of delivery delay and dissimilarity cost is minimized under prior knowledge of request rates. To this end, we apply integer relaxation and we tackle a minimax primal-dual formulation of the relaxation problem via a variant of gradient descent ascent, namely the Hybrid Block Successive Approximation (HiBSA) introduced in \cite{9}. HiBSA is known to converge to a stationary point of the relaxed minimax problem \cite{9}. Moreover, for the scenario in which the request rates are unknown, we present an online stochastic version of the algorithm that adapts to the requests observed over time.

The rest of this paper is organized as follows. In Sec. \RNum{2}, we describe the network model. The offline optimization problem is formulated and addressed in Sec. \RNum{3}. In Sec. \RNum{4}, we consider the online version of the similarity-based caching problem, and introduce the online scheme. Numerical results are presented in Sec. \RNum{5}. Finally, we offer some conclusions in Sec. \RNum{6}.

\vspace{-5pt}

\section{System Model}

As illustrated in Fig.1, we consider a network $\mathcal{N}=\{\mathcal{V}, \mathcal{E}\}$ consisting of a set of nodes $\mathcal{V}$ and a set $\mathcal{E}$ of undirected transmission links between pairs of nodes. The network delivers contents from a given set of popular contents $\mathcal{F}$ to devices connected to one of the network nodes. For every content $f\in\mathcal{F}$, there is a subset $\mathcal{S}_f\in\mathcal{V}$ of nodes, referred to as source nodes, that permanently store content $f$. We denote as $\tau_{uv}$ the average delivery delay between nodes $u,v\in\mathcal{V}$ and $(u,v)\in \mathcal{E}$ for any content in $\mathcal{F}$. A request consists of a content $f\in\mathcal{F}$ and of a fixed path $p$ through the network. Path $p$ is an acyclic sequence of nodes $p=\{p_{1},...,p_{|p|}\}$, where $p_1$ is the node receiving the request, $p_{k}\in\mathcal{V}$ for all edges $(p_{k},p_{k+1})\in\mathcal{E}$ with $k=1,2,...,|p|-1$, and $p_{|p|}\in\mathcal{S}_f$. Note that there may exist distinct requests $(f,p)$ and $(f,p')$ with paths $p$ and $p'$ sharing an arbitrary subsets of nodes. Once a request $(f,p)$ is received by the network, a request for content $f$ is routed through $p$ until a suitable content is found and delivered, through the same path, to the requesting device. Unlike \cite{10}, in which the request must be satisfied by delivering the requested content $f$, here we allow for similarity-based delivery \cite{4}: A content $f'\in\mathcal{F}$ different from the requested one may be delivered as long as the content selection satisfies a desirable trade-off between delivery latency and content similarity.

A similarity matrix of nonnegative values describes the similarity between pairs of contents in $\mathcal{F}$. Similarity may account for properties such as language, authors, genres, and so on. In this paper, following \cite{4}, we define a dissimilarity matrix $D = [d(f,f')]_{f,f'\in\mathcal{F}} \in {\mathbb{R}^{+}}^{|\mathcal{F}|\times|\mathcal{F}|}$, where $d(f,f')\in \mathbb{R}^{+}$ denotes the cost of delivering content $f'$ when the requested content is $f$. Naturally, we have $d(f,f)=0$ for all contents $f\in \mathcal{F}$.

The caching policy of node $v$ is defined by a vector $x_v=[x_{v,f}]_{f\in\mathcal{F}}$, where $x_{vf}\in\{0,1\}$ indicates whether node $v\in \mathcal{V}$ stores content $f\in\mathcal{F}$: we have $x_{vf}=1$ if node $v$ stores content $f$ and otherwise we set $x_{v,f}=0$. The overall caching policy of the network is defined by the matrix $X=[x_v]_{v\in\mathcal{V}}$. Due to cache capacity and source node cache constraints, we have the inequalities
\begin{align}
	&\sum\limits_{\substack{f\in \mathcal{F},\\ v\notin\mathcal{S}_f}} x_{v,f} \leq C_v, \quad \mbox{for all}\ v \in \mathcal{V}. \\
	&\mbox{and } x_{v,f} = 1, \mbox{for all } v\in\mathcal{S}_f.
\end{align}

Let $\mathcal{R}$ denote the set of requests $(f,p)$ that can be received by the network. Instances of requests $(f,p)\in\mathcal{R}$ are received according to independent Poisson processes, with arrival rate $\lambda_{(f,p)}$ (requests/s) for request $(f,p)$. We are interested in optimizing the caching decision matrix $X$, along with the delivery decision matrix $Q = [q_{(f,p),f'}]_{(f,p)\in\mathcal{R},f'\in\mathcal{F}}\in \{0,1\}^{|\mathcal{R}|\times|\mathcal{F}|}$. Variable $q_{(f,p),f'} \in\{0,1\}$ indicates the network decision to deliver content $f'$ for request $(f,p)$: We set $q_{(f,p),f'}=1$ if the network delivers content $f'$ in lieu of the requested $f$ through path $p$, and otherwise we have $q_{(f,p),f'}=0$. Only contents $f'$ that are cached or permanently stored at nodes $v\in\{p_1,...p_{|p|-1}\}$ can be selected for delivery for request $(f,p)$. Accordingly, we set $q_{(f,p),f'}=0$ if $x_{v,f'}=0$ for all $v\in p$. This constraint can be expressed as
\begin{align}
	& q_{(f,p),f'} \left( \prod_{k=1}^{|p|} \left( 1- x_{p_{k},f'} \right) \right) \leq 0, \nonumber \\
	& \qquad \mbox{for all } (f,p)\in\mathcal{R} \mbox{ and } f'\in \mathcal{F}.
\end{align}
\noindent Moreover, only one content is selected to be delivered for each request $(f,p)$, which can be expressed as
\begin{equation}
	\sum\limits_{f'\in\mathcal{F}} q_{(f,p),f'} = 1, \quad \mbox{for all}\ (f,p) \in \mathcal{R}.
\end{equation}

\vspace{-5pt}

\section{Offline Optimization}

In this section, we study the problem of minimizing the weighted sum of average delivery latency and dissimilarity cost with respect to the caching decision matrix $X=[x_{v,f}]_{v\in\mathcal{V},f\in\mathcal{F}}$ and the similarity-based delivery decision matrix $Q=[q_{(f,p),f'}]_{(f,p)\in\mathcal{R},f'\in\mathcal{F}}$. We assume here that the arrival rate matrix $\lambda = [ \lambda_{(f,p)}]_{(f,p)\in\mathcal{R}}$ is known, and that the problem is solved offline before the runtime delivery phase. 

\subsection{Problem Formulation}

As in \cite{10}, we assume that the delay of delivering the response message is much larger than that of forwarding a request. Therefore, the delay of delivering content $f'$ for request $(f,p)$ for a given caching matrix $X$ is written as
\begin{equation}
    t_{(f,p),f'}(X) = \sum_{k=1}^{|p|-1} \tau_{p_{k+1}, p_k} \prod_{k'=1}^k \left( 1 - x_{p_{k'},f'} \right). 
\end{equation}

\noindent We consider the problem of minimizing the weighted sum of average delay (5) plus the weighted sum of average dissimilarity cost under constraints (1)-(4). The cost of delivering content $f'$ for request $(f,p)$ consists of delay $t_{(f,p),f'}(X)$ and dissimilarity cost, written as $c_{(f,p),f'}(X)=t_{(f,p),f'}(X) + \alpha d(f,f')$, and parameter $\alpha$ is a nonnegative constant that quantifies the penalty in terms of the latency cost that is incurred by one unit of dissimilarity cost. The optimization problem is defined as the minimization
\begin{subequations}\label{orig opt}
	\begin{align}
		\min\limits_{X,Q} &\ c(X,Q)\triangleq \sum_{(f,p)\in\mathcal{R}} \lambda_{(f,p)} \sum_{f'\in\mathcal{F}}  q_{(f,p),f'} c_{(f,p),f'}(X) \label{orig obj} \\ 
		\text{s.t.} \ & X\in\{0,1\}^{|\mathcal{V}|\times|\mathcal{F}|}, Q\in\{0,1\}^{|\mathcal{R}|\times|\mathcal{F}|}, \\
		& \sum\limits_{f\in \mathcal{F}, v\notin\mathcal{S}_f} x_{v,f} \leq C_v, \ \mbox{for all}\ v \in \mathcal{V}, \\
		& x_{v,f}=1, \ \mbox{for all } v\in\mathcal{S}_f, \\
		& \sum_{f\in\mathcal{F}}q_{(f,p),f'}=1, \ \mbox{for all } (f,p)\in\mathcal{R}, \\
		& h_{(f,p),f'}(X,Q) \leq 0,\ \mbox{for all } (f,p)\in\mathcal{R},f'\in \mathcal{F} \label{nonconvex constraint},
	\end{align}
\end{subequations}

\noindent where we define $h_{(f,p),f'}(X,Q) \triangleq q_{(f,p),f'} ( \prod_{k=1}^{|p|} ( 1- x_{p_{k},f'} ) )$. All constraints in \eqref{orig opt} is introduced before. Reference \cite{10} studied the special case of problem \eqref{orig opt} in which $\alpha \rightarrow \infty$. Under this assumption, the only feasible solution for matrix $Q$ is given as $q_{(f,p),f'}=0$ for all $f'\neq f$ and $q_{(f,p),f}=1$, i.e., content $f$ is delivered for any request $(f,p)$. Therefore, the optimization is only over the caching matrix $X$.

In the following subsections, we tackle problem \eqref{orig opt} through the following steps: \textit{(i)} the integer constraints on matrices $X$ and $Q$ are relaxed, and a minimax formulation is introduced; \textit{(ii)} a variant of the gradient descent ascent algorithm, namely HiBSA \cite{9}, is applied to define the iterative procedure that converges to a stationary point of the minimax problem; \textit{(iii)} and a greedy rounding method is applied to obtain integer solutions for variables $X$ and $Q$.

\subsection{Integer Relaxation and Problem Reformulation}

In order to address the problem \eqref{orig opt}, we first relax the binary variables in matrices $X$ and $Q$ to lie in the interval $[0,1]$. The relaxed problem is still non-convex on account of the objective function \eqref{orig obj} and the constraint \eqref{nonconvex constraint}. We proceed by defining the Lagrangian function
\begin{equation}
    L(X,Q,\mu) = c(X,Q)+ h(X,Q,\mu), 
\end{equation}

\vspace{-3pt}
\noindent where we wrote
$h(X,Q,\mu) = \sum_{(f,p)\in\mathcal{R}}\lambda_{(f,p)} \sum_{f'\in\mathcal{F}} \mu_{(f,p),f'} h_{(f,p),f'}(X,Q) $, and $\mu=[\mu_{(f,p),f'}]_{(f,p)\in\mathcal{R},f'\in\mathcal{F}}$ are the Lagrangian multipliers for constraint \eqref{nonconvex constraint}. The multiplication by the requests' rate $\lambda_{(f,p)}$ is introduced in $h(X,Q,\mu)$ in order to simplify the online design presented in Sec. \RNum{4}. We then consider the problem
\begin{align}\label{minimax opt}
	&\min\limits_{X,Q}\max\limits_{\mu} \quad L(X,Q,\mu) \nonumber \\
	\text{s.t.} \ & X\in[0,1]^{|\mathcal{V}|\times|\mathcal{F}|}, Q\in[0,1]^{|\mathcal{R}|\times|\mathcal{F}|},(6c)-(6e).
\end{align}

\vspace{-3pt}
\noindent Note that the optimal solution $(X,Q)$ of problem \eqref{minimax opt} coincides with that of the mentioned relaxation of problem \eqref{orig opt} \cite[Chapter 5]{11}. The problem \eqref{minimax opt} is a nonconvex-concave minimax optimization problem, which is non-convex in the primal variables $X$ and $Q$ and concave (affine) in the dual variables $\mu$.

\vspace{-5pt}

\subsection{A Variant of the Gradient Descent Ascent Algorithm}

\begin{algorithm}[t]
\caption{HiBSA Algorithm with Rounding}
\label{alg1}
\begin{algorithmic}[1]
\STATE \textbf{Input:} $S(0)\in \Omega_S,\mu(0)\in\Omega_{\Psi}$; $\eta_s,\eta_{\mu},\{\gamma(n)\}$
\STATE \textbf{Output:} $S=(X,Q)$
\REPEAT
\STATE Compute $S(n+1) = \mathcal{P}_{\Omega_S}( S(n)-\eta_s\nabla_{S}L(n))$
\STATE Compute $\mu(n+1) = \big( \big(1+\gamma(n)\eta_{\mu}  \big) \mu(n) + \eta_{\mu}\nabla_{\mu}L(S(n+1),\mu(n))  \big)^{+}$
\UNTIL{Stopping criterion is satisfied}
\STATE Each node $v\in\mathcal{V}$ allocates cache to the uncached content with the largest $x_{v,f}$ until no cache is available
\STATE For each request $(f,p)\in\mathcal{R}$, set $q_{(f,p),f'}=1$ for $f'=\arg\max_{f''\in\mathcal{F}}q_{(f,p),f''}$, s.t. $h_{(f,p),f''}\leq 0$
\end{algorithmic}
\end{algorithm}

To tackle problem \eqref{minimax opt}, we apply the HiBSA algorithm introduced in \cite{9}, which leverages gradient descent for the minimization over primal variables $S=(X,Q)$ and gradient ascent for the maximization problem of dual variables $\mu$. The HiBSA algorithm is proved in \cite{9} to converge to a stationary solution by solving a sequence of convex minimization problem and concave maximization problem with suitable regularization terms. To apply this scheme, we need to first identify strongly convex and concave approximation functions for the primal variables $S$ and dual variables $\mu$, respectively, that satisfy the conditions in \cite{9}.

Let the $S(n) = (X(n),Q(n))$ and $\mu(n)$ denote the $n$-th iterate of the primal variables and dual variables. Since the Lagrangian function $L(S,\mu)$ is twice-differentiable, it has an $l_s$-Lipschitz constant with respect to primal variables $S$, i.e., the largest eigenvalue of the Hessian matrix of $L(S,\mu)$. For approximation function $\bar{L}(S;S(n),\mu(n))$ in the primal variables $S$, we consider the function $\bar{L}(S;S(n),\mu(n))=L(S(n),\mu(n))+\nabla_S L(S(n),\mu(n))^{\mathrm{T}}(S-S(n))+\frac{1}{2\eta_s} \lVert S - S(n) \rVert^2$ for some constant $\eta_s\leq 1/l_s$. Since the Lagrangian function $L(S,\mu)$ is linear in $\mu$, the approximation function for dual variable $\mu$ can be directly defined as $\bar{L}(\mu;S(n+1),\mu(n))=L(\mu,S(n+1))-\frac{1}{2\eta_{\mu}} \lVert \mu - \mu(n) \rVert^2$ for some constant $\eta_{\mu}>0$. The introduced approximation functions satisfy Assumptions in \cite{9}, since they are respectively strongly convex and concave; they provide respectively upper bound and lower bound for Lagrangian function at the current iterate; they guarantee gradient consistency; and they have Lipschitz continuous gradients. The convex minimization problem for primal variable $S$ solved at iteration $n$ by HiBSA is defined as
\begin{align}\label{primal iterative}
    & S(n+1) = \arg\min_{\Omega_S} (\bar{L}(S;S(n),\mu(n))) \nonumber \\
    & = \mathcal{P}_{\Omega_S}( S(n)-\eta_s\nabla_{S}L(n)),
\end{align}

\noindent where $\nabla_{S}L(n) = \{\{ \frac{\partial L(n)}{\partial x_{v,f'}}\} , \{ \frac{\partial L(n)}{\partial q_{(f,p),f'}}\}\}$ denote the overall gradients for the primal variables, and $\mathcal{P}_{\Omega_S}$ is the projection to the convex subset $\Omega_S$ of primal variables defined by constraints (6c)-(6e) \cite[Chapter 8]{11}. Let function $k_v(p)$ return the position of node $v$ in path $p$, so that we have $k_v(p)\in\{1,...,|p| \}$ if $v\in p$ and $k_v(p)=-1$ otherwise. The derivatives of $L(S,\mu)$ evaluated at the $n$-th iterate can be computed as
\begin{align}\label{gradient x}
    &\frac{\partial L(n)}{\partial x_{v,f'}}  = - \sum\limits_{(f,p)\in\mathcal{R}}\lambda_{(f,p)} \Bigg [q_{(f,p),f'}(n)  \nonumber \\
    & \cdot \Bigg( \sum_{k=1}^{|p|-1} \tau_{p_{k+1}, p_{k}} \prod_{\substack{k'=1\\ 1\leq k_v(p) \leq k \\ k' \neq k_v(p)}}^k \big( 1 - x_{p_{k'},f'}(n) \big)\Bigg) \nonumber \\
    & - \sum\limits_{(f,p)\in\mathcal{R}}\mu_{(f,p),f'}(n)q_{(f,p),f'}(n)  \nonumber \\
    & \cdot \Bigg( \prod_{\substack{k=1 \\ k_v(p)\geq 1 \\ k \neq k_v(p) }}^{|p|} \left( 1- x_{p_{k,f},f'}(n) \right) \Bigg)\Bigg],
\end{align}
\noindent and
\begin{align}\label{gradient q}
    &\frac{\partial L(n)}{\partial q_{(f,p),f'}} = \lambda_{(f,p)}\Bigg[\Bigg( t_{(f,p),f'}(X(n)) + \alpha d(f,f')\Bigg) \nonumber \\
    & + \mu_{(f,p),f'}(n) \Bigg( \prod_{k=1}^{|p|} ( 1- x_{p_{k},f'}(n)) \Bigg)\Bigg].
\end{align}

\noindent Similarly, the concave maximization problem for dual variables $\mu$ at iteration $n$ is defined as
\begin{align}\label{dual iterative}
    & \mu(n+1) = \arg\max_{\mu\in\mathbb{R}^{+}}\big( \bar{L}(\mu;S(n+1),\mu(n)) - \frac{\gamma(n)}{2}\lVert \mu(n) \rVert^2 \big) \nonumber \\
    & = \Big( \big(1+\gamma(n)\eta_{\mu}  \big) \mu(n) + \eta_{\mu}\nabla_{\mu}L(n)  \Big)^{+},
\end{align}

\noindent where $\gamma(n) = \frac{1}{\eta_{\mu} n^{1/4}}$ is a perturbation parameter, which satisfy Assumption C in \cite{9}. In \eqref{dual iterative}, we have function $(x)^{+}=x$ if $x\geq0$ and $(x)^{+}=0$ otherwise. The gradients with respect to dual variable $\mu$ at iteration $n$ are denoted as $\nabla_{\mu}L(n) = \{\frac{\partial L(n)}{\partial \mu_{(f,p),f'}}\}$, and are computed as
\begin{align}\label{gradient mu}
    &\frac{\partial L(n)}{\partial \mu_{(f,p),f'}}  \nonumber \\
    & =  \lambda_{(f,p),f'}\Big[q_{(f,p),f'}(n+1) \Big( \prod_{k=1}^{|p|} \left( 1- x_{p_{k},f'}(n+1) \right) \Big)\Big].
\end{align}

The overall HiBSA algorithm is summarized in Algorithm 1. The stopping criterion is given as $|L(n+1)-L(n))|\leq \delta$, where $\delta>0$ is the desired accuracy, and we now discuss how to perform rounding.

\subsection{Rounding Method}

In order to obtain an integer solution for the output of the HiBSA algorithm, a greedy rounding algorithm is applied to round first the caching decision matrix $X$ and then the delivery decision matrix $Q$. Considering the first step of rounding $X$, each node $v\in\mathcal{V}$ selects contents to cache by following the order of decreasing values of $x_{v,f}$ while the cache capacity constraint is met with equality. Then, in a similar manner, for each request $(f,p)\in\mathcal{R}$, the delivery decision is set to $q_{(f,p),f'}=1$ for content $f'=\arg\max_{f''\in\mathcal{F}}q_{(f,p),f''}$, s.t. $h_{(f,p),f''}\leq 0$ for all $f''\in \mathcal{F}$. Note that quantifying the loss due to quantization remains an open problem, which does not seem tractable with standard tools such as those used in \cite{10}.

\section{Online Optimization}

In this section, we consider the scenario in which the arrival rate matrix $\lambda$ is a priori unknown. The requests are received according to the independent Poisson processes described in Sec. \RNum{2}. In order to tackle the problem of optimizing caching and delivering decision matrices $(X,Q)$ in this scenario, we introduce an online algorithm to solve the minimax problem \eqref{minimax opt}. Following the offline solution presented in Sec. \RNum{3}, the algorithm leverages stochastic gradient descent for primal variables and stochastic gradient ascent for dual variables. Moreover, an online greedy rounding method is used to determine the cache allocation at nodes.

Time is partitioned into periods of equal length $T>0$. In time period $t$, the number of instances of each request $(f,p)\in\mathcal{R}$ is a Poisson variable with mean $\lambda_{(f,p)}T$. Denote as $\mathcal{R}_t$ the multi-set of requests received in the $t$-th time slot. Note that a request $(f,p)$ may appear multiple times in $\mathcal{R}_t$. For each received request $(f,p)$, the network delivers the content $f'$ that satisfies $f'=\arg\max_{f''\in\mathcal{F}}q_{(f,p),f''}, \mbox{ s.t. } h_{(f,p),f''}\leq 0$. This means that we choose the content $f'$ with the largest value of the current delivery decision variable $q_{(f,p),f'}$ subject to cache availability. We denote as $\mathcal{R}_t'$ the multi-set of triples $(f,p,f')$ containing request $(f,p)\in\mathcal{R}_t$ and associated delivered content $f'$. 

Since the only measured delays are for requests $(f,p,f')\in\mathcal{R}_t'$, the derivatives \eqref{gradient x}, \eqref{gradient q} and \eqref{gradient mu} are computed only for variables $x_{v,f'}$ and $q_{(f,p),f'}$ with $(f,p,f')\in\mathcal{R}_t'$. A stochastic estimate $\hat{\partial} L(n)/\partial x_{v,f'}$ for the derivative in \eqref{gradient x} can be specifically obtained as $\sum_{(f,p,f')\in\mathcal{R}_t'}A/T$, where $A$ is the term in the square bracket in \eqref{gradient x}. In a similar way, stochastic estimates for the derivatives in \eqref{gradient q} and \eqref{gradient mu} can be obtained by choosing $A$ as the terms in the square bracket in \eqref{gradient q} and \eqref{gradient mu}, respectively. Following the same arguments as in \cite[Lemma 1]{10}, these stochastic derivatives are unbiased estimates of the true derivatives and they have finite variance. At the end of any time slot $t$, estimates of the derivatives in \eqref{gradient x}, \eqref{gradient q} and \eqref{gradient mu} are computed as discussed above and applied using steps 4 and 5 in Algorithm 1 by replacing the gradients $\nabla_S L(n)$ and $\nabla_{\mu} L(n)$ with the discussed stochastic estimates. These two steps are followed by greedy rounding as for steps 7 and 8 of Algorithm~\ref{alg1}.

\begin{figure}
    \centering
    \includegraphics[width=0.4\textwidth]{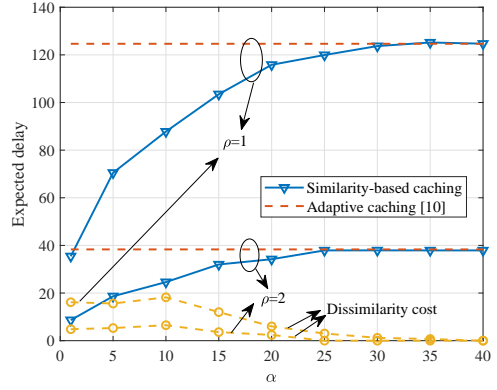}
    \caption{Expected delay of similarity-based caching compared with adaptive caching [10], along with the dissimilarity cost, for two values of the content popularity $\rho$ obtained by similarity-based caching.}
    \label{fig2}
\end{figure}

\section{Numerical Experiments}

In this section, we provide numerical results concerning a grid-2D network topology with $|\mathcal{V}|=25$ nodes and $|\mathcal{E}|=100$ edges \cite{10}. The average delay $\tau_{uv}$ over an edge $(u,v)$ follows an uniform distribution in the interval $[1,10]$. Each node has a caching capacity $C_v=2$ contents. The total number of contents in $\mathcal{F}$ is 10, and for each $f\in\mathcal{F}$, a node $v\in\mathcal{V}$ is randomly selected as the source node that permanently stores content $f$. The set $\mathcal{R}$ of requests, with cardinality $|\mathcal{R}|=40$, is generated as follows. We first select a subset $\mathcal{V}_s\in \mathcal{V}$ of nodes with $|\mathcal{V}_s|=12$ that can generate requests. For each request $(f,p)$, a content $f$ is selected from $\mathcal{F}$ following a Zipf distribution with parameter $\rho>0$; and the forwarding path $p$ is selected as the shortest path from a randomly selected starting node in $\mathcal{V}_s$ to the source node of the requested $f$. With set $\mathcal{R}$ fixed, we set $\lambda_{(f,p)}=1$ as the arrival rate for every $(f,p)\in \mathcal{R}$. The dissimilarity of content $f$ and $f'$ is modeled as $d(f,f') = |f-f'|^{\beta}$, where $\beta$ is a non-negative constant. We set $\beta=3$. The performance metrics is measured as the expected delay of requests, i.e., $D(X,Q) = \sum_{(f,p)\in\mathcal{R}}\lambda_{(f,p)}\sum_{f'\in\mathcal{F}}q_{(f,p),f'}t_{(f,p),f'}$. We compare the obtained performance with the offline algorithm introduced in \cite{10}, which does not enable similarity caching and is referred to as \emph{adaptive caching}. We will also provide a comparison with the state-of-art \emph{per-cache} scheme $q$LRU-$\Delta C$. This scheme always delivers the most similar content in the cache of the starting node for each request.

\begin{figure}
    \centering
    \includegraphics[width=0.4\textwidth]{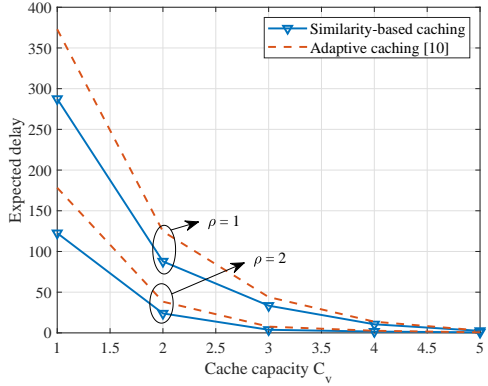}
    \caption{Expected delay of similarity-based caching scheme versus the cache capacity $C_v$ of all nodes compared with the adaptive caching scheme in \cite{10}.}
    \label{fig3}
\end{figure}

First, we evaluate the offline algorithm in Algorithm 1, referred to as \emph{similarity-based caching}, as a function of the weight $\alpha$ given to the dissimilarity cost in (6a) for $C_v=2$ at all nodes. We set $\eta_s=10^{-3}$ and $\eta_{\mu}=1$. In Fig. \ref{fig2}, when $\alpha$ is small, \emph{similarity-based caching} is seen to obtain a significantly lower expected delay as compared with \emph{adaptive caching} by delivering similar contents instead of the requested contents. As $\alpha$ grows larger, delivering different contents is increasingly penalized, and the performance converges to that of \emph{adaptive caching} \cite{10}. Fig. \ref{fig2} also shows that the dissimilarity cost of similarity-based caching decreases with $\alpha$. We also observe the more significant gains obtained by \emph{similarity-based caching} when $\rho$ is larger, corresponding to a request distribution more concentrated around the most popular contents.

Fig. 3 shows the expected delay performance versus the cache capacity $C_v$, assumed to be equal for all nodes. Here, we set $\alpha=10$. It is observed that, as cache resources become abundant, the two schemes obtain similar results, while \emph{similarity-based caching} is better able to use limited caching resources for the given value of $\alpha$.  

\begin{figure}
    \centering
    \includegraphics[width=0.4\textwidth]{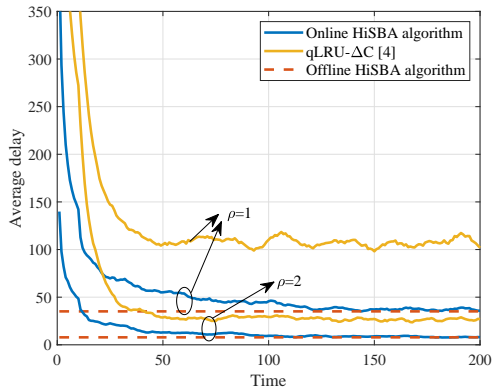}
    \caption{Average delay for the proposed online similarity-based caching scheme and $q$LRU-$\Delta C$ as a function of the number of time slots. The proposed online performance converges to the average delay obtained by the offline scheme.}
    \label{fig4}
\end{figure}

Finally, we evaluate the performance of the online HiBSA algorithm. To this end, we simulate the request (Poisson) processes and plot the average delay obtained with the current iterates $(X,Q)$, as a function of the number of time slots. The length of time period is $T=1$. We set the step size for updating variable $X$ as $\eta_{x}=1\times10^{-3}$, the step size for updating variable $Q$ as $\eta_q=1\times10^{-4}$, and $\eta_{\mu}=1$. Note that we have found it useful to set the step size $\eta_{x}$ to be larger than $\eta_q$, which suggests that the change in $Q$ should be more gradual than for $X$. The delay is averaged in a window comprising the last ten time slots. The plot corresponds to one realization of the request processes. It is also seen that the online HiBSA algorithm can significantly outperform $q$LRU-$\Delta C$ thanks to network-wide coordination. We also observe that online similarity-based caching scheme approaches the performance of the offline scheme as more requests are processed. Convergence is particularly fast for more concentrated popularity distributions, i.e., for larger $\rho$. This is because in this case it is sufficient to optimize the caching delivery decision variables only for the more popular contents in order to reap most of the benefits of caching.

\section{Conclusions}

In this work, we have studied a multi-hop caching network in which similarity-based delivery is allowed. Both offline and online optimization of caching and delivery policy have been considered. The proposed solutions are based on a variant of gradient descent ascent that minimizes the weighted sum of delay and dissimilarity cost of the requests. Interesting future directions are integrating the use of advanced wireless edge caching strategies \cite{3}, and implementing larger-scale networks.


\begin{thebibliography}{15}

\bibitem{1}
Y. Li, H. Xie, Y. Wen, C. Chow and Z. Zhang, ``How Much to Coordinate? Optimizing In-Network Caching in Content-Centric Networks," \emph{IEEE Trans. on Netw, Service Manag.}, vol. 12, no. 3, pp. 420-434, Sep. 2015.

\bibitem{2}
S. Borst, V. Gupta and A. Walid, ``Distributed Caching Algorithms for Content Distribution Networks," in \emph{2010 Proc. IEEE INFOCOM}, USA, Mar. 2010, pp. 1-9.

\bibitem{3}
S. M. Azimi, O. Simeone, A. Sengupta and R. Tandon, ``Online Edge Caching and Wireless Delivery in Fog-Aided Networks With Dynamic Content Popularity," \emph{IEEE J. Sel. Areas Commun.}, vol. 36, no. 6, pp. 1189-1202, June 2018.


\bibitem{4}
M. Garetto, E. Leonardi and G. Neglia, ``Similarity Caching: Theory and Algorithms," in \emph{2020 Proc. IEEEE INFOCOM}, China, April 2020.


\bibitem{5}
D. Zhang, J. Wang, D. Cai, and J. Lu, ``Self-taught hashing for
fast similarity search," in \emph{Proc. 33rd Int. ACM SIGIR Conf. Res. Development Inf. Retrieval}, July 2010, pp. 18–25.

\bibitem{6}
S. Pandey, A. Z. Broder, F. Chierichetti, V. Josifovski, R. Kumar, and S. Vassilvitskii, ``Nearest-neighbor caching for content-match applications,” in \emph{Proc. 18th Int. Conf. World Wide Web}, April 2009, pp. 441–450.

\bibitem{7}
P. Sermpezis, T. Giannakas, T. Spyropoulos and L. Vigneri, ``Soft Cache Hits: Improving Performance Through Recommendation and Delivery of Related Content," \emph{IEEE J. Sel. Areas Commun.}, vol. 36, no. 6, pp. 1300-1313, June 2018.

\bibitem{8}
P. Sermpezis, T. Spyropoulos, L. Vigneri, and T. Giannakas, ``Femtocaching with soft cache hits: Improving performance with related content
recommendation,” in \emph{2017 Proc. IEEE GLOBECOM}, Singapore, Dec. 2017, pp. 1–7.




\bibitem{9}
S. Lu, I. Tsaknakis, M. Hong, and Y. Chen, ``Hybrid Block Successive Approximation for One-Sided Non-Convex Min-Max Problems: Algorithms and Applications,'' to appear on \emph{IEEE Trans. Signal Processing}.


\bibitem{10}
S. Ioannidis, and E. Yeh, ``Adaptive Caching Networks With Optimality Guarantees," \emph{IEEE/ACM Trans. Netw.} , vol. 26, no. 2, pp. 737-750, April 2018.



\bibitem{11}
S. Boyd, S. P. Boyd, and L. Vandenberghe, \emph{Convex optimization}. Cambridge university press, 2004.


	
\end{thebibliography}
\end{document}